\documentclass[12pt]{article}

\usepackage[font=small, labelfont=bf]{caption}
\usepackage{graphicx}
\usepackage{float}
\usepackage[T1]{fontenc}
\usepackage{braket}
\usepackage{bm}
\usepackage{hyperref}
\usepackage[numbers,sort&compress]{natbib}
\usepackage{makecell}
\usepackage{textcomp}
\usepackage{multirow}
\usepackage{booktabs}
\usepackage{authblk}
\usepackage{amsmath}

\usepackage{times}
\usepackage[margin=2.5cm]{geometry}
\usepackage{setspace}
\newcommand{\diag}{\mathrm{diag}}

\begin{document}

\title{Towards ultracompact photonic chips using higher-order modes in closely spaced waveguides}

\author[1]{F. Yousry\thanks{Author to whom any correspondence should be addressed. Email: \href{mailto:fahmy.ahmed@aalto.fi}{fahmy.ahmed@aalto.fi}}}
\author[1]{P. Hildén}
\author[1]{R. Kolkowski}
\author[1]{A. Shevchenko}

\affil[1]{Department of Applied Physics, Aalto University, PO Box 13500, FI-00076, Aalto, Finland}

\date{} % Optional: remove date

\maketitle

\begin{abstract}
    Photonic integrated circuits are gaining traction in the field of telecommunications and information processing for their low-loss and high-throughput data transmission in comparison to electronic integrated circuits. However, they are still not used as widely as their electronic counterparts due to a relatively large footprint of photonic chips. One limiting factor to their size is the need to separate optical components by distances on the order of the working wavelength or larger to minimize optical crosstalk between them. In this work, we consider the fundamental and higher-order modes in closely spaced straight and bent waveguides with relatively small cross sections and find that higher-order modes allow one to substantially reduce the crosstalk in both cases. This can be used to considerably reduce the dimensions of photonic chips. We also propose on-chip components that allow selective excitation of higher-order modes. In addition, we design a directional coupler, a 3-dB splitter, and a Mach-Zehnder interferometer capable of operating on higher-order modes. Other ultracompact photonic-chip components, such as optical interconnects, switches, transceivers, and phased waveguide arrays for on-chip LiDAR scanners, can be designed as well based on similar principles.
\end{abstract}


\section{Introduction}

Photonic integrated circuits (PICs) hold great promise for complementing their electronic counterparts, since they exhibit lower transmission losses and larger bandwidths \cite{karabchevskyOnchipNanophotonicsFuture2020,koenderinkNanophotonicsShrinkingLightbased2015,nisaEfficientOnChipCommunication2024}. However, integration of PICs with nanofabricated electronic devices has been hindered by their inherently large sizes dictated by crosstalk between photonic components. Typically, the separation of crosstalk-free PIC components is larger than the working free-space wavelength ($\lambda$), i.e., it is huge in comparison to the achievable nm-scale separation of nanoelectronic elements \cite{karabchevskyOnchipNanophotonicsFuture2020}. High-refractive-index materials, such as silicon, offer better confinement of optical fields in the waveguides, which can be used to reduce the PIC size \cite{shekharRoadmappingNextGeneration2024,thomsonRoadmapSiliconPhotonics2016a}. However, decreasing the cross section of the waveguide core makes the field less confined in the core and more extended in the cladding, so the crosstalk increases again \cite{zhaoUltrahighSensitivityMachZehnder2021}.

Crosstalk reduction between closely spaced waveguides has been demonstrated, e.g., in waveguide superlattices and bent waveguide arrays \cite{gatdulaGuidingLightBent2019,zhouArtificialGaugeField2023,zafarObandTETMmode2023,songHighdensityWaveguideSuperlattices2015,hildenMatrixAnalysisHighdensity2024}. More recently, it has been shown that crosstalk can be completely suppressed between straight waveguides with a separation distance on the order of $\lambda/10$ by making use of higher-order modes \cite{mauryaCrosstalkReductionClosely2022}. In fact, higher-order modes in photonic waveguides have been considered also previously as possible additional channels of information that can be processed independently using mode division multiplexing (MDM) \cite{liMultimodeSiliconPhotonics2019,wuModeDivisionMultiplexingSilicon2017}. Silicon waveguides are the preferred platform for MDM systems because, in addition to a high-index core and low losses in the telecommunication wavelength range, they exhibit strong mode dispersion, which is highly valued in multimode nanophotonic structures \cite{liMultimodeSiliconPhotonics2019}. Many silicon-based devices required for MDM systems have been developed, such as on-chip mode converters \cite{gonzalez-andradeUltraBroadbandModeConverter2018,xuModeConversionTrimming2024}, multiplexers and demultiplexers \cite{kawaguchiModeMultiplexingDemultiplexing2002,liCompactTwomodeDemultiplexer2014,gonzalez-andradeUltraBroadbandModeConverter2018}, multimode waveguide crossings \cite{xuDualmodeWaveguideCrossing2016}, and chip-fiber couplers \cite{laiEfficientSpotSize2017}. These devices are usually designed to operate on higher-order transverse electric or magnetic modes TE$_{l0}$ and TM$_{l0}$, where $l$ corresponds to the order of the field along the horizontal direction (parallel to the plane of the substrate). However, crosstalk suppression has been shown to be possible only for out-of-plane higher-order modes TE$_{01}$ and TM$_{01}$, which can be called the azimuthally and radially polarized modes, respectively \cite{mauryaCrosstalkReductionClosely2022}.

In this work, we demonstrate the possibility of creating high-density PICs that operate on higher-order modes by taking advantage of their crosstalk suppression capabilities. Section 2 of the manuscript presents a study of crosstalk in straight and bent waveguide arrays carrying either higher-order or fundamental modes. We show that higher-order modes can exhibit considerably lower crosstalk in both straight and bent arrays. In section 3, we introduce two approaches for selectively exciting specific higher-order modes in waveguides with simultaneously reduced core size and crosstalk. Section 4 describes the design of basic building blocks of PICs, including a directional coupler, a 3-dB splitter, and a Mach–Zehnder interferometer for higher-order azimuthally polarized modes. The last section summarizes the results of this work.

\section{Crosstalk suppression via higher-order modes}

\begin{figure}
	\includegraphics[width=\textwidth]{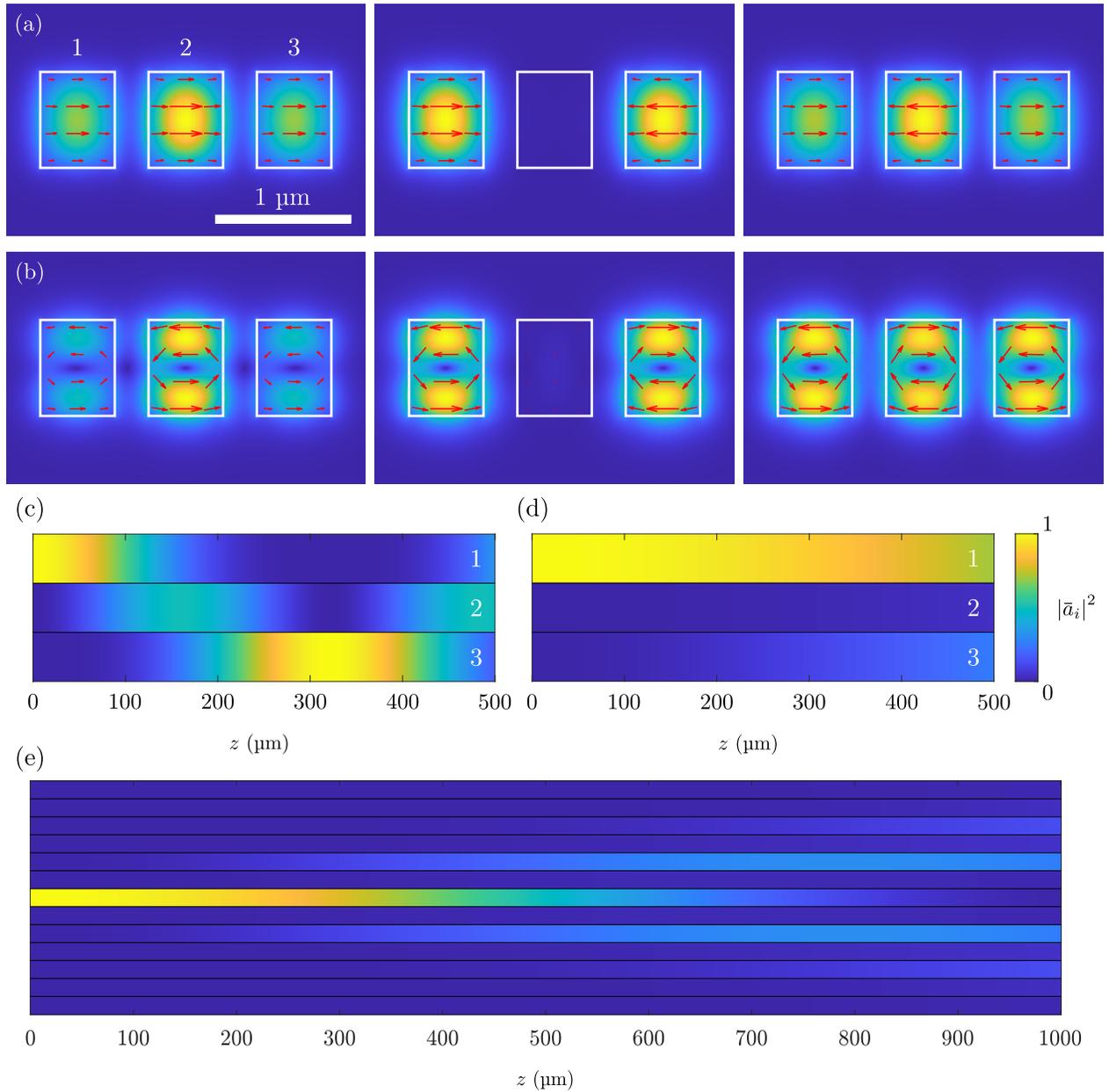}
	\caption{The field distributions of (a) TE and (b) AZ supermodes of three coupled waveguides. The arrows show the direction and relative strength of the electric field. The normalized optical power as a function of propagation distance $z$ in the three waveguides, shown for (c) TE mode and (d) AZ mode excitation in waveguide 1. In (e), the propagation of the normalized power originally coupled to the central waveguide of a 13-waveguide array is illustrated. The power is seen to skip the nearest waveguides when spreading in the array.}
	\label{fig:TE_vs_AZ_str_array}
\end{figure}

Consider $N$ identical waveguides forming an array with period $\Lambda$. Let them be parallel to the $z$-axis and periodically repeated along the $x$-direction. Suppose that one of the guided modes of an individual waveguide (e.g., TE$_{00}$ mode) is excited in the $i^{\text{th}}$ waveguide with an amplitude $\bar{a}_i(z=0)$. Here, $\bar{a}_ i(z)$ is the longitudinally varying amplitude of the electric field $\bar{a}_i(z)\bar{\mathbf{e}}_i(x,y)$,  where $\bar{\mathbf{e}}_i(x,y)$ is the normalized transverse distribution of the mode field. In this work, the electric and magnetic field distributions, denoted by $\mathbf{e}$ and $\mathbf{h}$, respectively, are normalized such that the power flow along the direction of propagation is 1 W, i.e., 
\begin{equation}
    \frac{1}{2} \int\int_{-\infty}^{\:\infty} \text{Re}\{\mathbf{e}\times\mathbf{h}^*\}\cdot\hat{\mathbf{z}}\,dx\,dy = 1 \; \text{W}.
    \label{eq:normalization}
\end{equation}
The bar above $\bar{a}_i$ designates that the mode belongs to an individual waveguide and not to the array. For an array of waveguides, the amplitude of a supermode will be marked with a tilde, $\tilde{a}_i$. Being subject to crosstalk, the amplitude of the mode propagating in the $i^{\text{th}}$ waveguide, $\bar{a}_i(z)$, satisfies the differential equation \cite{hardyCoupledModesMultiwaveguide1986}
\begin{equation}
    -\text{i} \frac{\partial\bar{a}_i}{\partial z} = (\bar{\beta}+\delta_i) \bar{a}_i + \kappa(\bar{a}_{i-1}+\bar{a}_{i+1})+\chi(\bar{a}_{i-2}+\bar{a}_{i+2})+...,
	\label{eq:diffEq}
\end{equation}
where $\bar{\beta}$ is the propagation constant of the mode in an isolated waveguide, $\delta_i$ is the self-coupling coefficient, $\kappa$ is the coupling coefficient of the mode to the nearest waveguides, and $\chi$ is the coupling coefficient to the second-nearest waveguides. Further coupling terms can be neglected, if $\Lambda$ is sufficiently large. The phase accumulation rate of $\bar{a}_i$ is increased by $\delta_i$ compared to an isolated waveguide, because the neighboring waveguides increase the effective refractive index of the cladding and consequently, also the effective propagation constant of the mode \cite{shevelevaLocalfieldEffectivebackgroundEffects2023}. We can assume that coefficients $\delta_i$ are the same for all the waveguides ($\delta_i=\delta$) except for the waveguides at the edges of the array, as they have neighboring waveguides only on one side, and hence, $\delta_i=\delta/2$ for them \cite{cooperNumericallyassistedCoupledmodeTheory2009}. Often, the coupling coefficients for waveguides beyond the nearest neighbors (such as $\chi$) can be neglected (see, e.g., \cite{kaponSupermodeAnalysisPhaselocked1984,xiaSupermodesCoupledMultiCore2016}). However, this is not the case for higher-order modes with suppressed crosstalk, as will be shown below. Due to the coupling, all amplitudes $\bar{a}_i(z)$ depend on the amplitudes of the surrounding waveguides, each satisfying a similar differential equation. Together, these equations form a linear system of differential equations that can be written as \cite{kaponSupermodeAnalysisPhaselocked1984,xiaSupermodesCoupledMultiCore2016,hardyCoupledModesMultiwaveguide1986}
\begin{equation}
	-\text{i}\frac{\partial\bar{\mathbf{a}}}{\partial z} = \mathbf{H}\bar{\mathbf{a}},
	\label{eq:Hamiltonian}
\end{equation}
using vector $\bar{\mathbf{a}} = [\bar{a}_1\ \bar{a}_2\ ...\ \bar{a}_N]^{T}$ and matrix
\begin{equation}
    \mathbf{H} = \bar{\beta}\mathbf{I} + 
    \left[
    \begin{array}{ccccccc}
    \frac{\delta}{2} & \kappa & \chi & \cdots & & & \\
    \kappa & \delta & \kappa & \chi & \cdots & & \\
    \chi & \kappa & \delta & \kappa & \chi & \cdots & \\
    \vdots & \ddots & \ddots & \ddots & \ddots & \ddots & \\
    & & \chi & \kappa & \delta & \kappa & \chi \\
    & & & \chi & \kappa & \delta & \kappa \\
    & & & & \chi & \kappa & \frac{\delta}{2}
    \end{array}
    \right]
    \label{eq:azH}
\end{equation}
The superscript $T$ denotes transpose. The solution to (\ref{eq:Hamiltonian}) is determined by the eigenvalues $\tilde{\bm{\beta}} = [\tilde{\beta}_1 \ \tilde{\beta}_2 \ ... \tilde{\beta}_N]^{T}$ and eigenvectors $\tilde{\bar{\mathbf{c}}}_j$ of matrix $\mathbf{H}$. The eigenvectors form the matrix $\tilde{\bar{\mathbf{C}}} = [\tilde{\bar{\mathbf{c}}}_1 \ \tilde{\bar{\mathbf{c}}}_2 \ ...\ \tilde{\bar{\mathbf{c}}}_N]^{T}$. As a function of propagation distance $z$, the solution is \cite{greenbergOrdinaryDifferentialEquations2012}
\begin{equation}
		\bar{\mathbf{a}}(z)= \tilde{\bar{\mathbf{C}}}^{-1} \diag\left[\exp(\text{i}\tilde{\bm{\beta}}z)\right] \tilde{\bar{\mathbf{C}}} \bar{\mathbf{a}}(0)
	\label{eq:prop_str_wga}.
\end{equation}
The normalized eigenvectors $\tilde{\bar{\mathbf{c}}}_j$ describe the eigenmodes of the $entire$ array, with the $i^{\text{th}}$ element describing the relative amplitude of light in the $i^{\text{th}}$ waveguide. The corresponding eigenvalues, $\tilde{\beta}_j$, are the propagation constants of the eigenmodes, also known as supermodes. Function $\diag(\cdot)$ constructs a diagonal matrix from the input vector. Examples of supermodes for a three-waveguide array are shown in figure \ref{fig:TE_vs_AZ_str_array} for (a) the fundamental TE$_{00}$ modes (called simply TE modes for brevity) and (b) the azimuthally polarized higher-order TE$_{01}$ modes (that we call AZ modes). The supermodes have been calculated using COMSOL Multiphysics (mode-analysis eigenvalue
solver of the Wave Optics module). The waveguides are made of Si ($n=3.48$) embedded in glass ($n=1.44$), forming an array with period $\Lambda=800$ nm. The cross section of each waveguide has the vertical and horizontal sizes of 710 and 550 nm, respectively. The free-space wavelength was set to 1550 nm. At this wavelength, an individual waveguide can host ten modes: four TE modes, five TM modes, and one hybrid TE-TM mode. Note that, essentially, the AZ mode, when disturbed, can be coupled only to the other three TE modes. This coupling can appear due to imperfections in the waveguide structure or as a result of too tight bending of the waveguides. In general, a waveguide array can be designed to exhibit significantly reduced crosstalk for higher-order modes by optimizing its geometrical parameters (height, width, and period). This can be done by maximizing the figure of merit introduced in reference \cite{mauryaCrosstalkReductionClosely2022}. For demonstration purposes, we will optimize the height while keeping the other parameters unchanged.

The crosstalk-induced spread of light in a waveguide array described by (\ref{eq:prop_str_wga}) is analogous to optical diffraction in free space described by Fourier optics. Given the amplitudes in individual waveguides in $\bar{\mathbf{a}}$,  the amplitude of the $j^{\text{th}}$ supermode, $\tilde{a}_j$, is equal to the product $\tilde{\bar{\mathbf{c}}}_j^{T}\bar{\mathbf{a}}$. Moreover, the matrix transformation
\begin{equation}
	\tilde{\mathbf{a}} = \tilde{\bar{\mathbf{C}}}\bar{\mathbf{a}}
\end{equation}
represents an arbitrary field propagating in the waveguide array in terms of the supermodes rather than the fields in individual waveguides. In the supermode basis, the propagation of the guided field is governed simply by the phase change of each supermode $\phi_j=\tilde{\beta}_jz$. Due to differences in the values of propagation constants $\tilde{\beta}_j$, the supermodes accumulate different phases upon propagation and interfere differently, leading to a different distribution of the amplitudes across the individual waveguides. To recover the amplitudes in the individual waveguides, we apply the inverse transformation $\tilde{\bar{\mathbf{C}}}^{-1}=\bar{\tilde{\mathbf{C}}}=\tilde{\bar{\mathbf{C}}}^{T}$, yielding
\begin{equation}
\bar{\mathbf{a}}(z) = \tilde{\bar{\mathbf{C}}}^{-1}\tilde{\mathbf{a}}(z).
\end{equation}

Considering the TE and AZ supermodes shown in figures \ref{fig:TE_vs_AZ_str_array}(a) and \ref{fig:TE_vs_AZ_str_array}(b), we find that the propagation constants of the AZ supermodes are closer to each other, corresponding to lower crosstalk. The lower crosstalk of the AZ modes can be seen in figures \ref{fig:TE_vs_AZ_str_array}(c) and \ref{fig:TE_vs_AZ_str_array}(d), where the normalized optical power in each waveguide, $|\bar{a}_i(z)|^2$, is shown for both the TE and AZ modes excited in the first waveguide of the array. Ultimately, in a pair of waveguides, the AZ supermodes can be made completely degenerate ($\tilde{\beta}_1=\tilde{\beta}_2$), meaning that the coupling between the waveguides is completely suppressed ($\kappa=0$) \cite{mauryaCrosstalkReductionClosely2022}. However, if the number of waveguides is increased, light can start to exhibit crosstalk between more distant waveguides. For example, in figure \ref{fig:TE_vs_AZ_str_array}(d), light is seen to be coupled from the first to the third waveguide, rather than the second one. There are two reasons for that. The first reason comes from the fact that the crosstalk between the first and second waveguides is still suppressed, and hence, we have $|\chi|>|\kappa|$. This is clearly illustrated in figure \ref{fig:TE_vs_AZ_str_array}(e), where the AZ mode is excited in the central waveguide of a 13-waveguide array. As the optical power propagates through the array, it skips the nearest neighboring waveguide and couples to the next one, slowly removing the power from the original waveguide. The second reason is that the propagation constants in the first and third waveguides are smaller than in the second waveguide due to the difference in the self-coupling coefficients (\textit{cf.} crosstalk reduction in waveguide superlattices \cite{songHighdensityWaveguideSuperlattices2015}). This is in contrast with the usual case of $\left|\kappa\right|\gg\left|\chi\right|$, where $\mathbf{H}$ simplifies to a tridiagonal Toeplitz matrix, resulting in supermodes described by a discrete sine transformation matrix of type 1, i.e., $\tilde{\bar{c}}_{ji}=\sqrt{2/(N+1)}\sin(\pi ji/(N+1))$ and $\tilde{\beta}_j = \bar{\beta}+2\kappa\cos(\pi j/(N+1))$ \cite{kaponSupermodeAnalysisPhaselocked1984,xiaSupermodesCoupledMultiCore2016}. In figure \ref{fig:TE_vs_AZ_str_array}, the TE supermodes closely follow this simple model, while the AZ supermodes do not, making them more difficult for analytical treatment. Here, we find and analyze the higher-order supermodes numerically. For each calculated $j^{\text{th}}$ supermode, the elements of $\tilde{\bar{\mathbf{c}}}_j$ can be obtained from the overlap integrals between the supermode and the modes of single waveguides, as described in detail in \cite{hildenMatrixAnalysisHighdensity2024}.

To gain perspective on the space saved by using higher-order modes, we compare our optimized waveguides with optimized single-mode waveguides. The key parameter chosen for this comparative study is the center-to-center separation between two neighboring waveguides. We fix the coupling length of the single-mode waveguides at 3 mm to match that of the AZ mode in a pair of our optimized waveguides and find the required separation. Each optimized single-mode waveguide has a square cross section with the side length maximized while maintaining its single-mode operation. This ensures a minimized evanescent field coupling between the waveguides. We find that two waveguides with an optimized side length of 320 nm must be separated by 1.34 \textmu m. In this particular case, our approach gives a size reduction factor of about 0.6. Note, however, that in principle, AZ modes allow the crosstalk between two waveguides to be completely eliminated, rendering the coupling length infinite. To achieve the same with single-mode waveguides, one must place them at an infinitely large distance from each other.

\begin{figure}
	\centering
	\includegraphics[width=\linewidth]{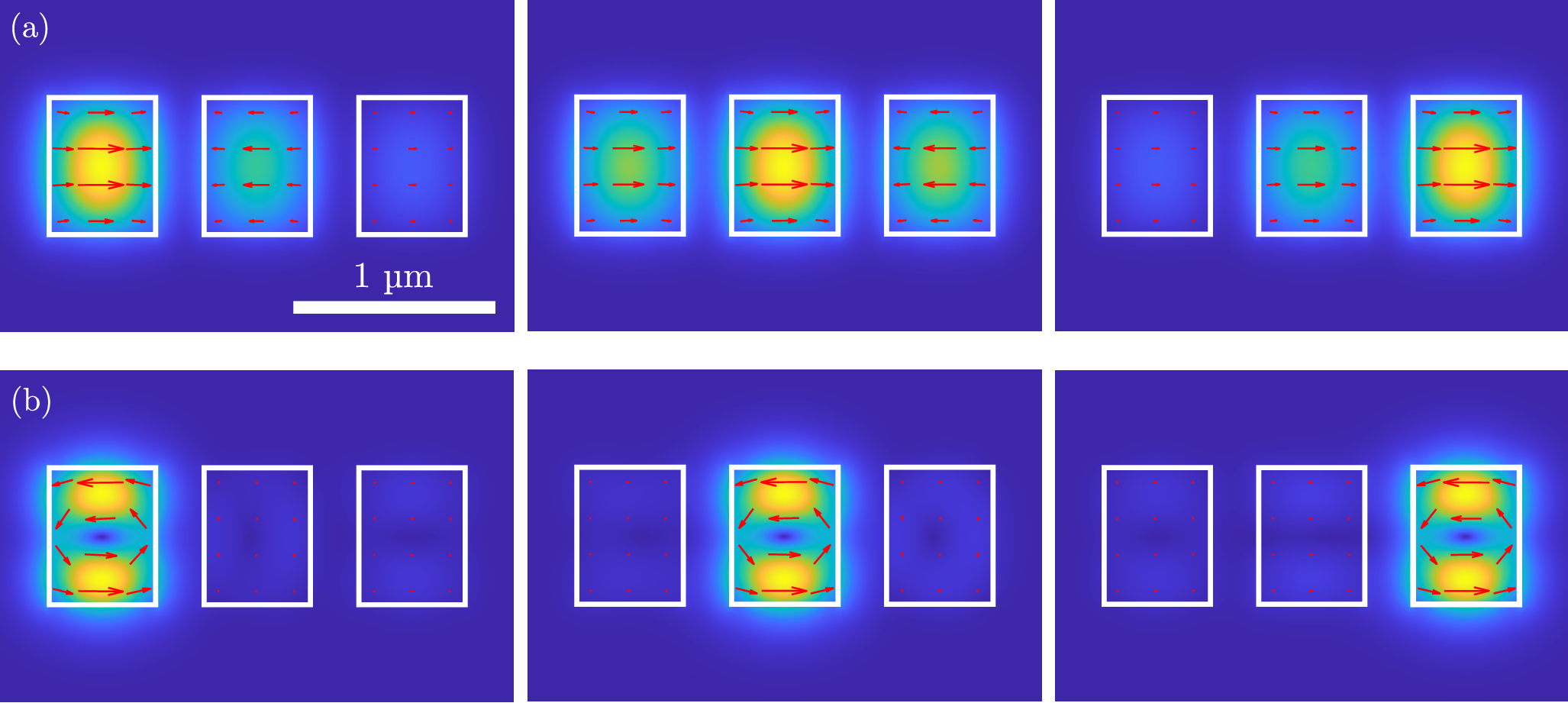}
	\caption{The field distributions of (a) TE and (b) AZ supermodes of a bent waveguide array with $R=1$ mm. The supermodes have been calculated numerically using the mode solver of the COMSOL Wave Optics module in a 2D axisymmetric space. }
	\label{fig:TE_vs_AZ_bent_supermodes}
\end{figure}

In addition to utilizing higher-order modes, bending the waveguide array offers an alternative method for suppressing crosstalk \cite{lenzBlochOscillationsArray1999a,zafarObandTETMmode2023,hildenMatrixAnalysisHighdensity2024}. The two approaches are complementary, meaning that they can be combined to improve the action of each other. This stands in contrast with crosstalk suppression via waveguide superlattices, which is counteracted by bending \cite{gatdulaGuidingLightBent2019}. In \cite{hildenMatrixAnalysisHighdensity2024},  the transformation matrix 
$\breve{\bar{\mathbf{C}}}$ has been introduced to change the basis between the field amplitudes in individual waveguides and the supermodes of a bent waveguide array:
\begin{equation}
	\breve{\mathbf{a}} = \breve{\bar{\mathbf{C}}}\bar{\mathbf{a}}.
	\label{eq:str_to_bent}
\end{equation}
Here, the elements of vector $\breve{\mathbf{a}}$ are the amplitudes of the supermodes in the bent waveguide array that remain invariant under the angular propagation of light in the array. The transformation matrix $\breve{\bar{\mathbf{C}}}$ is determined by the product \cite{hildenMatrixAnalysisHighdensity2024}
\begin{equation}
    \breve{\bar{\mathbf{C}}} = \breve{\tilde{\mathbf{C}}} \tilde{\bar{\mathbf{C}}},
\end{equation}
where, matrix $\breve{\tilde{\mathbf{C}}}$ corresponds to the transformation from the supermode basis of the straight waveguide array to that of the bent waveguide array. It is defined as $\breve{\tilde{\mathbf{C}}} = [\breve{\tilde{\mathbf{c}}}_1 \ \breve{\tilde{\mathbf{c}}}_2 \ ...\ \breve{\tilde{\mathbf{c}}}_N]^{T}$.  The vector $\breve{\tilde{\mathbf{c}}}_k$ is the $k^{\text{th}}$ eigenvector of matrix
\begin{equation}
    \mathbf{M} = \langle\tilde{\beta}\rangle\left[\left(R-\Lambda \frac{N+1}{2}\right)\mathbf{I} + \Lambda \left( \diag(\boldsymbol{\rho}) + \tilde{\bar{\mathbf{C}}} \diag(\mathbf{i}) \tilde{\bar{\mathbf{C}}}^{-1} \right) \right].
\end{equation}
Here, $\langle\tilde{\beta}\rangle$ is the mean value of the straight-waveguide-array supermode propagation constants approximately equal to $\bar{\beta}$, $R$ is the mean radius of curvature of the array, vector $\mathbf{i}$ is equal to $[1\;2\; ...\; N]^T$, and $\mathbf{I}$ is the identity matrix.  The elements of vector $\boldsymbol{\rho}$ are defined as \cite{hildenMatrixAnalysisHighdensity2024}
\begin{equation}
	\rho_j=\frac{R\Delta\tilde{\beta}_j}{\Lambda \langle\tilde{\beta}\rangle},
	\label{eq:rho}
\end{equation} 
where, $\Delta\tilde{\beta}_j=\tilde{\beta}_j-\langle\tilde{\beta}\rangle$ is the deviation from $\langle\tilde{\beta}\rangle$. Note that, for large values of $\boldsymbol{\rho}$, the matrix $\mathbf{M}$ is nearly diagonal, implying that $\breve{\tilde{\mathbf{C}}}=\mathbf{I}$, and hence, we have $\breve{\bar{\mathbf{C}}} = \tilde{\bar{\mathbf{C}}}$. This means that the supermodes of the bent waveguide array have the same profiles as in the corresponding straight waveguide array.  On the other hand, when the elements of $\boldsymbol{\rho}$ are small,  the matrix $\mathbf{M}$ can be diagonalized  to show that $\breve{\bar{\mathbf{C}}}=\mathbf{I}$. This implies that in waveguide arrays with a small radius of curvature, the supermodes are the modes confined in individual waveguides, with no crosstalk. Moreover, the expression for $\boldsymbol{\rho}$ in (\ref{eq:rho}) implies that, for AZ supermodes with small differences in propagation constants $\Delta\tilde{\beta}_j$, strong confinement can be achieved even if the radius of curvature is large. This is shown to be the case in figures \ref{fig:TE_vs_AZ_bent_supermodes}(a) and \ref{fig:TE_vs_AZ_bent_supermodes}(b), where the TE and AZ supermodes of a bent waveguide array are presented for $R=1$ mm. The ability to confine light in individual waveguides while maintaining a large radius of curvature helps mitigate possible bending-induced losses. Note that, at $R<30$ \textmu m, the phase mismatch between the AZ mode and other higher-order modes in neighboring waveguides can be compensated for by bending. This may lead to an increase of the intermodal coupling that can be significant at long propagation distances. However, this is not unique to AZ modes, and the fundamental TE modes at smaller bending radii also start to exhibit this type of coupling. Bent waveguide arrays can be used for on-chip LiDAR scanners \cite{heckHighlyIntegratedOptical2017}, in which (1) the waveguide period should be smaller than the wavelength to eliminate diffraction orders and (2) the crosstalk between the waveguides should be absent to keep the phases in the waveguides independent. Using AZ modes would fulfill both the requirements.

\section{Excitation of a higher-order AZ mode}

Although higher-order modes offer clear advantages in terms of crosstalk suppression and PIC miniaturization, it is challenging to excite a single higher-order mode in a multimode waveguide. In this section, we introduce two approaches to selectively excite the AZ mode in a waveguide with the same dimensions as in section 2. These dimensions allow only the two lowest-order TE modes to be excited, the TE and AZ modes. The first of the approaches presented below is easier from a fabrication point of view, while the second one offers a high coupling efficiency.

\begin{figure}
    \centering
    \includegraphics[width=\linewidth]{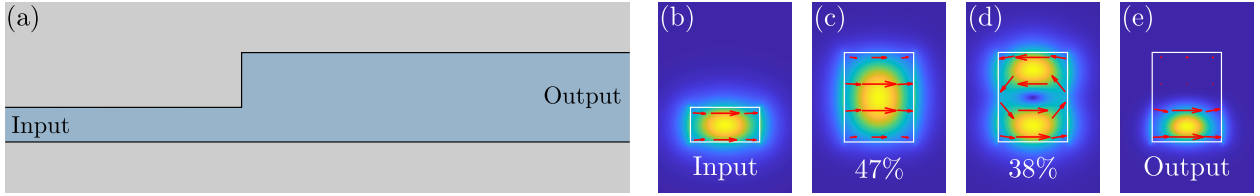}
    \caption{(a) The side view of a waveguide coupler used to excite a higher-order AZ mode in the output waveguide. The electric field distribution of (b) the fundamental TE mode of the thinner input waveguide and (c) the TE and (d) the AZ modes excited at the interface. The total output field at the interface is shown in (e).}
    \label{fig:step_coupler}
\end{figure}

The first approach is based on making the input part of the waveguide thinner, as shown in figure \ref{fig:step_coupler}(a). Light is initially coupled to the fundamental TE mode of the thinner part. The field distribution in this mode, for the height of the waveguide equal to 275 nm, is shown in figure \ref{fig:step_coupler}(b). When the mode enters the thicker part of the waveguide, two eigenmodes of the thicker part are excited. They are the fundamental TE mode, shown in figure \ref{fig:step_coupler}(c), and the AZ mode, shown in figure \ref{fig:step_coupler}(d). The excitation efficiencies of the two modes are calculated using the mode overlap integral \cite{okamotoFundamentalsOpticalWaveguides2006,paszkiewiczApproximationMethodFast2024}
\begin{equation}
    \Gamma_i =  \left\| \;\frac{1}{4}\int\int_{-\infty}^{\:\infty} (\bar{\mathbf{e}}_i^*\times\bar{\mathbf{h}}_{\text{input}}+\bar{\mathbf{e}}_{\text{input}}\times\bar{\mathbf{h}}_i^*)\cdot\hat{\mathbf{z}}\,dx\,dy\;\right\|^2,
\end{equation}
where $\Gamma_i$ corresponds to the excitation efficiency of the $i^\text{th}$ mode of the output waveguide, $\bar{\mathbf{e}}_{\text{input}}$ and $\bar{\mathbf{h}}_{\text{input}}$ are the electric and magnetic fields of the TE mode of the thin waveguide, and $\bar{\mathbf{e}}_i$ and $\bar{\mathbf{h}}_i$ are the field distributions of the $i^\text{th}$ mode of the output waveguide. Recall that the fields are normalized in accordance with (\ref{eq:normalization}). The height of the input waveguide is optimized to maximize the overlap integral of the input TE mode with the output AZ mode. The efficiencies of coupling to the TE and AZ modes of the output waveguide are 47 \% and 38 \%, respectively.  The total field of the two excited modes is shown in figure \ref{fig:step_coupler}(e). This profile is not exactly the same as the input-field profile, which is attributed to a small loss of power to other, presumably leaky modes of the system. To filter out the TE mode and obtain a pure AZ-mode excitation, the low crosstalk of the AZ mode can be used. The filter includes a piece of another waveguide separated from the first waveguide by 250 nm. At this distance, the crosstalk between the AZ modes is suppressed, while the TE mode will be totally coupled to the auxiliary waveguide in a propagation distance of 230 \textmu m. In this way, the TE mode can be removed. The theoretical excitation efficiency of the AZ mode with this method is therefore equal to 38 \%. The filter is similar to a standard on-chip directional coupler designed for fundamental TE modes. In principle, the length of the filter can be reduced by maximizing the difference between the propagation constants of the fundamental (TE) supermodes, while keeping the difference of the propagation constants of the AZ supermodes as small as possible. Another way to filter out the TE mode is to exploit the fact that the AZ mode is dark in the center of the waveguide cross section, while the TE mode is bright. Hence, adding a thin layer of a highly absorbing material at half the height of the waveguide can quickly remove the fundamental mode by absorption. We want to emphasize that such a filter may be used only once or twice in a PIC, as it is needed only at the input and maybe also at the output of the device, in which case its size reduction is not very critical.

\begin{figure}
    \centering
    \includegraphics[width=\linewidth]{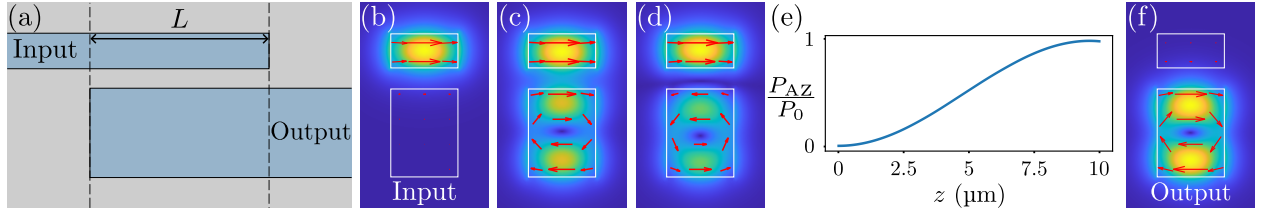}
    \caption{(a) The side view of a two-waveguide coupler used to excite an AZ mode in the output waveguide. The length of the coupler is $L$. The input field distribution is shown in (b). The two supermodes of the two-waveguide system, combining the TE mode of the thin waveguide and the AZ mode of the thick waveguide, are shown in (c) and (d). The blue line in (e) represents the fraction of the power coupled to the AZ mode as the field propagates through the structure. The output field distribution is shown in (f).}
    \label{fig:vertical_coupler}
\end{figure}

The second approach is based on a waveguide coupler, in which a thin waveguide is positioned above the waveguide designed to carry the AZ mode, as shown in figure \ref{fig:vertical_coupler}(a). This arrangement converts the input field (shown in figure \ref{fig:vertical_coupler}(b)) into a superposition of two supermodes of the two-waveguide system. Each supermode is a combination of the TE mode of the thin waveguide with the AZ mode of the thick waveguide (see figures \ref{fig:vertical_coupler}(c) and \ref{fig:vertical_coupler}(d)). In this structure, the width of the thin waveguide is taken to be equal to that of the thick waveguide, while its height and separation from the thick waveguide are optimized for the two supermodes excited in the coupler to have equal amplitudes. Note that the AZ modes in the two supermodes oscillate out of phase at the entrance of the coupler, making it possible to realize a superposition in which the field profile resembles the field of the thin waveguide (see figure \ref{fig:vertical_coupler}(b)).
As the field propagates farther in the coupler, the phase difference between the supermodes increases. If the length of the coupler is selected such that, at its end, the phase difference is equal to $\pi$, the power is fully transferred to the AZ mode of the lower waveguide.  In figure \ref{fig:vertical_coupler}(e), we show the fraction of the power transferred to the AZ mode as a function of the field propagation distance in the coupler. If the length of the coupler, $L$, is chosen to be 10 \textmu m, the coupling efficiency to the AZ mode can reach 98 \%, as obtained in our numerical simulations. In this design, the height of the thin waveguide and its surface-to-surface separation from the thick waveguide are 275 and 170 nm, respectively. At these parameters, coupling of light to other modes is negligible due to large phase mismatch.

Both approaches can be used to selectively excite the AZ mode in a waveguide. Once excited, it can be coupled to other waveguides or guided through a waveguide array with suppressed crosstalk. In such case, the crosstalk will depend solely on the waveguide separations. 

\section{On-chip components for AZ modes}

\begin{figure}
    \centering
    \includegraphics[width=\linewidth]{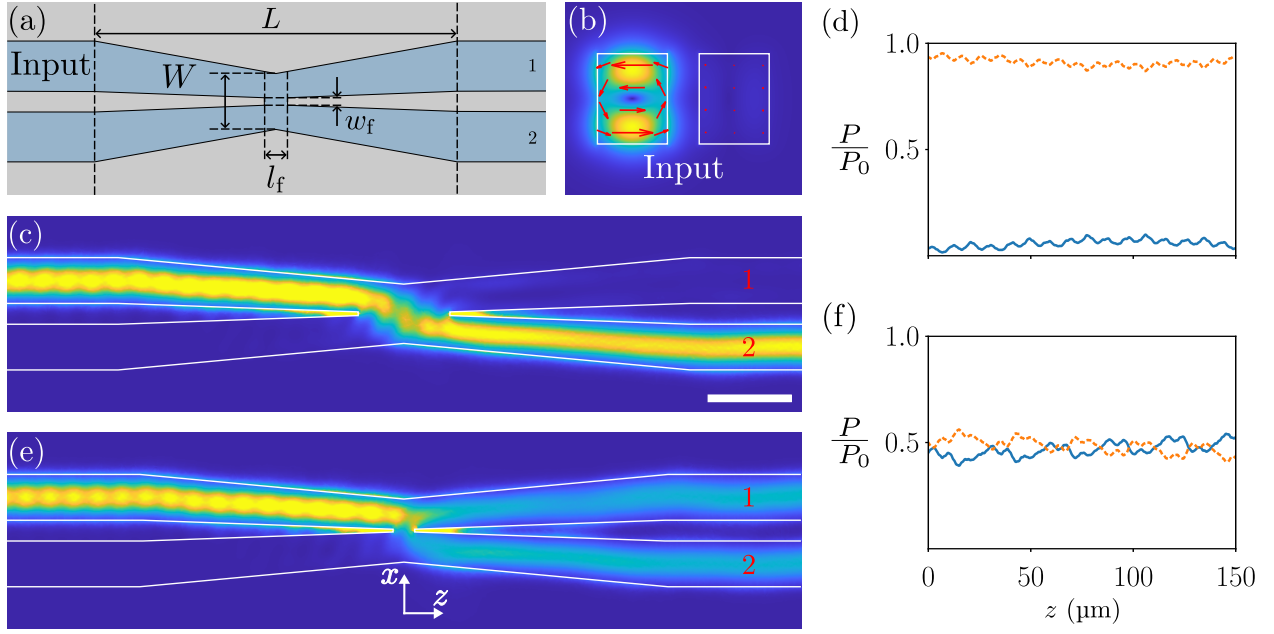}
    \caption{(a) Top view of the proposed geometry for a directional coupler and a 3-dB splitter. (b) The field distribution at the input. The intensity distribution in the designed (c) directional coupler and (e) 3-dB splitter, normalized to the peak intensity at the input. The intensity distributions are shown for the $xz$-plane at $y=223$ nm measured from the center of the structure. The white scale bar in (c) corresponds to 1 \textmu m. In (d) and (f), the optical powers in waveguide 1 (blue line) and waveguide 2 (orange dashed line) are shown as functions of propagation distance $z$ from the output of the directional coupler and 3-dB splitter, respectively.}
    \label{fig:DC_splitter}
\end{figure}

Crosstalk is frequently used in on-chip directional couplers and power splitters. These devices are traditionally made of two waveguides that are locally positioned close to each other to increase crosstalk between their guided modes. The length of the coupler is optimized to obtain the desired power splitting between the waveguides. In the aforementioned waveguide arrays designed for AZ modes, the separation of the waveguides is already very small, 250 nm, and at the same time, the crosstalk is fully suppressed. Hence, an alternative approach to the design of a directional coupler or a 3-dB splitter must be found.  The structure we propose is based on tapering and fusing the two input waveguides together, as shown in figure \ref{fig:DC_splitter}(a). The structure is similar to a fused fiber coupler \cite{lianApplicationFusedTapering2025}. In contrast to conventional couplers, the two output waveguides do not need to be spaced apart (and, in fact, should not be) by more than 250 nm, because this spacing ensures their crosstalk-free operation.

To optimize the geometric parameters of the structure and calculate the percentage of the initial power coupled to each waveguide at the output, we use full-field simulations based on the COMSOL Multiphysics software, following the tutorial and an example provided by the COMSOL support team \cite{CopyingReusingBoundary,DirectionalCoupler}. The input and output boundaries of the structure are set at numerical ports. Each input and output pair of ports corresponds to one supermode of the two waveguides. The number of supermodes (port pairs) is denoted by $Q$. The scattering boundary condition is assigned to all other boundaries to absorb the fields scattered away from the structure. In the simulations, the structure is excited with the AZ mode in the left waveguide as shown in figure \ref{fig:DC_splitter}(b) (top waveguide in figure \ref{fig:DC_splitter}(a)). The COMSOL solutions are given in the form of the complex amplitudes of the supermodes of the two-waveguide system at the output, which can then be used to construct the total output electric and magnetic fields, $\mathbf{E}_\text{out}$ and $\mathbf{H}_\text{out}$, as follows:
\begin{equation}
    \mathbf{E}_\text{out} = \tilde{\mathbf{a}}^T \tilde{\mathbf{E}}, \; \mathbf{H}_\text{out} = \tilde{\mathbf{a}}^T \tilde{\mathbf{H}}.
\end{equation}
Here, we have $\tilde{\mathbf{E}} = [\tilde{\mathbf{e}}_1\ \tilde{\mathbf{e}}_2\ ...\ \tilde{\mathbf{e}}_Q]^{T}$ and $\tilde{\mathbf{H}} = [\tilde{\mathbf{h}}_1\ \tilde{\mathbf{h}}_2\ ...\ \tilde{\mathbf{h}}_Q]^{T}$, where each pair of elements $\tilde{\mathbf{e}}_j$ and $\tilde{\mathbf{h}}_j$ corresponds to the transverse electric and magnetic field distributions, respectively, of the $j^{\text{th}}$ supermode. The fields of the supermodes are normalized such that the total power flow along the propagation direction is 1 W (see (\ref{eq:normalization})). The vector $\tilde{\mathbf{a}} = [\tilde{a}_1\ \tilde{a}_2\ ...\ \tilde{a}_Q]^{T}$ contains the amplitudes of the supermodes at the output of the structure. Using the total output fields, one can then calculate the percentage of the power in each waveguide at the output. Ideally, the number of supermodes, $Q$, would be equal to two, corresponding to the two AZ supermodes of the array. However, some weak excitations of other modes can appear in the output field due to the non-adiabatic nature of the coupler. To study the effect of these parasitic modes on the output, the propagation of the field in the system after the transmission through the structure has been calculated as well. The propagated electric field at any $z$ is found as
\begin{equation}
    \mathbf{E}_\text{out}(z) = \tilde{\mathbf{a}}^T \diag\left[\exp( \text{i}\tilde{\bm{\beta}}z)\right] \tilde{\mathbf{E}}.
    \label{eq:prop_in_2wg}
\end{equation}
A similar formula can be written for the output magnetic field, $\mathbf{H}_\text{out}$. For a directional coupler, all the optical power should be transferred to the second waveguide. The values of the optimized geometric parameters $L$, $W$, $l_f$, and $w_f$ are found to be 6.9 \textmu m, 0.71 \textmu m, 1.1 \textmu m, and 43 nm, respectively. The normalized intensity distribution in the device in the $xz-$plane at $y=223$ nm is shown in figure \ref{fig:DC_splitter}(c). This plane is chosen to pass through one of the intensity peaks of the mode. The optical power is seen to be transferred from the first to the second waveguide. A weak standing-wave pattern seen in the left half of the device is due to a non-negligible reflection from the waveguide junction. The fraction of the initial power in each output waveguide as a function of the field propagation distance is shown in figure \ref{fig:DC_splitter}(d). A small fraction of the initial power still stays in the first waveguide (see the blue line), some power is lost upon reflection from and scattering at the waveguide junction (ca. 0.26 \%),  and about 10 \% of the power is coupled to the parasitic modes at the output. The presence of parasitic modes in the output field is manifested by small changes of the powers in the waveguides when the field propagates. These modes obviously have pronounced crosstalk.

For a 3-dB splitter, the parameters $L$, $W$, $l_f$, and $w_f$ of the structure are 6.36 \textmu m, 0.76 \textmu m, 270 nm, and 33 nm, respectively. As can be seen in figure \ref{fig:DC_splitter}(e), the device distributes the power evenly between the two waveguides. figure \ref{fig:DC_splitter}(f) shows the dependence of the power in the output waveguides on the propagation distance. Some parasitic modes (8.7 \% of the power) are also excited in this structure, leading to the rapid transfer of small fractions of the power between the waveguides. The total reflected power in this device is 0.38 \%. It is worth noting that the dimension $w_f$ does not require a nm-scale fabrication precision, because slight changes of the specified value (say, in the interval between 30 and 40 nm) have an insignificant effect on the device performance.

\begin{table}[ht]
    \centering
    \caption{A comparison of the directional coupler designed in this work with examples from the literature.}
    \label{table:comparison}
    \begin{tabular}{lllll}
        \toprule
        Device & This work & This work & \cite{bangerterInverseDesign2x22023} & \cite{piggottFabricationconstrainedNanophotonicInverse2017} \\
        \midrule
        Length (\textmu m) & 6.9 & 6.36 & 4 & 2.45 \\
        Width (\textmu m) & 1.35 & 1.35 & 4 & 1.2 \\
        Insertion loss (dB) & 0.13 & 0.22 & 0.034 & 0.46 \\
        \makecell[l]{Output waveguide\\coupling length (mm)} & 3 & 3 & 1.1 & 1 or 0.008 \\
        \bottomrule
    \end{tabular}
\end{table}

Table \ref{table:comparison} compares the proposed directional couplers with other state-of-the-art miniaturized devices introduced in the literature. The device metrics, except for the coupling length of the input or output waveguides, are taken directly from references \cite{bangerterInverseDesign2x22023} and \cite{piggottFabricationconstrainedNanophotonicInverse2017} at the optimal operation wavelength. The waveguide coupling length was calculated numerically using the dimensions and material compositions of the devices mentioned in the publications and the following standard formula for the coupling length \cite{mauryaCrosstalkReductionClosely2022}:
\begin{equation}
    L_{\pi} = \frac{\lambda}{2\left|n_\text{a}-n_\text{s}\right|},
\end{equation}
where $\lambda$ is the free-space wavelength, while $n_\text{a}$ and $n_\text{s}$ are the effective mode indices of the asymmetric and symmetric supermodes, respectively. The proposed devices are of comparable footprint and performance. Moreover, if another similar waveguide is placed next to any of the two devices at a center-to-center separation distance of 800 nm from one of the input waveguides, the coupling between the waveguide and the device stays negligibly small. The same can be said about two closely spaced directional couplers. This makes the size reduction of complex optical chips easier. Our components are seen to have a comparable, if not smaller, size and a comparable, if not better, performance with the other state-of-the-art miniaturized on-chip devices.

The two components considered above can be further optimized, modified, and made tunable, e.g., by adjusting the phase shift between the output modes with the help of an external electric field (parts of the structure can be made of lithium niobate or another material exhibiting the Pockels effect). Moreover, if needed, the parasitic modes can be filtered out from the output waveguides, e.g., by coupling them to an auxiliary waveguide positioned at a distance of 250 nm from one of the output waveguides of the coupler.

\begin{figure}
    \centering
    \includegraphics[width=\linewidth]{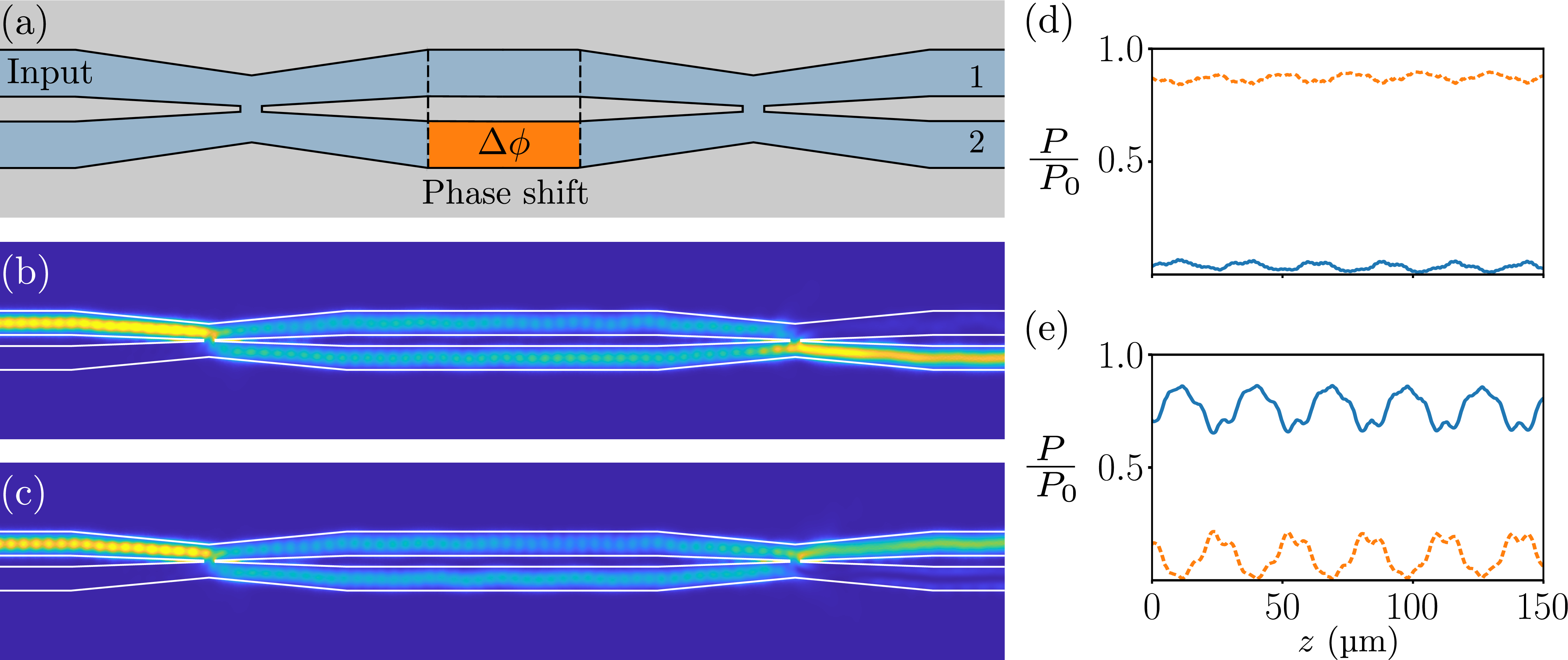}
    \caption{(a) Top view of an on-chip Mach-Zehnder interferometer with a phase shifting element introduced in one of the arms. The intensity distribution in the MZI for $\Delta\phi = 0$ and $\Delta\phi = \pi$ is shown in (b) and (c), respectively. The intensity is shown in the $xz$-plane normalized to its peak value in the input waveguide at $y=223$ nm. Optical powers in waveguide 1 (blue line) and waveguide 2 (orange dashed line) as functions of propagation distance $z$ after the MZI for (d) $\Delta\phi=0$ and (e) $\Delta\phi=\pi$.}
    \label{fig:MZI}
\end{figure}

As an application example, let us consider a Mach-Zehnder interferometer (MZI) designed for AZ modes. The device is constructed by cascading two 3-dB splitters as shown in figure \ref{fig:MZI}(a). The distance between the two splitters is 7 \textmu m. The input AZ mode is split into two modes by the first splitter. Then the modes propagate in the arms of the interferometer, where their phase difference ($\Delta\phi$) can be tuned, e.g., electro-optically. The second splitter combines the modes to propagate further in the output waveguides. In our simulations, we set $\Delta\phi$ either to 0 or to $\pi$. To induce the phase shift of $\pi$, we increase the effective refractive index of the mode in the straight part of one of the arms by $\Delta n =0.1$. This change introduces a phase mismatch between the AZ modes of the two waveguides, further suppressing crosstalk \cite{songHighdensityWaveguideSuperlattices2015}. In practice, the effective mode index can be adjusted all-optically or electrically. If the considered change $\Delta n=0.1$ is for some practical reason unachievable, one can increase the length of the arms by a desired factor, while reducing the required $\Delta n$ by the same factor. The increase in the total length of the MZI arms should have a negligible effect on the crosstalk between the waveguides, because the length of the arms is on the order of tens of micrometers, while the coupling length of the AZ modes in the arms is a few millimeters. In figure \ref{fig:MZI}, the intensity distributions in the MZI at $\Delta\phi$ equal to 0 and $\pi$ are shown in (b) and (c), respectively. In the first case, the power is completely transferred to the second waveguide, while in the second case the power mainly remains in the first waveguide, as expected. The numerically calculated reflected power is as low as 0.29 \% and 0.43 \% of the total power for $\Delta\phi$ equal to 0 and $\pi$, respectively. This power, however, is enough to lead to the standing-wave profile observed at the input. Indeed, the relative modulation amplitude of the standing-wave intensity is approximately equal to $2\sqrt{0.0043}$ for $\Delta\phi$ equal to $\pi$, which gives 13 \%.

We used (\ref{eq:prop_in_2wg}) to calculate the optical powers in the output waveguides as functions of the distance from the MZI. For $\Delta\phi = 0$ and $\Delta\phi=\pi$, the calculation results are shown in figures \ref{fig:MZI}(d) and \ref{fig:MZI}(e), respectively. As can be seen from figure \ref{fig:MZI}(d), most of the power is transferred to the second waveguide and only a small fraction of it is coupled back to the first waveguide, which is mostly due to the parasitic modes at the output (4.7 \% of the total power). In the case shown in figure \ref{fig:MZI}(e), the majority of the power remains in the initial waveguide. However, a certain non-negligible fraction of the power is seen to be coupled also to the second waveguide, exhibiting a pronounced periodic oscillation due to crosstalk associated with parasitic modes (20 \% of the total power). This unwanted effect should be minimized, e.g., by filtering these modes away from the output waveguides with the help of an auxiliary waveguide at a center-to-center separation of 800 nm from the output waveguide of interest. All the modes except the AZ mode will be coupled to the auxiliary waveguide and thus removed. The length of the filtering waveguide must be 29 \textmu m. The filter will remove more than 65 \% of the parasitic-mode power (13 \% of the total power) from the MZI output. The presented interferometer is an example of a PIC component that works reasonably well with AZ modes. One possible application of the MZI, in which parasitic modes are not filtered out, is an amplitude modulator for digital signals. In such a device, one can define a threshold intensity above which the signal is 1 and below which it is 0. In this case, the problem associated with the parasitic modes is absent. For interferometric applications, in which the total phase of the field is needed, the parasitic modes will decrease the accuracy of the measurements. However, for sensing and detection applications, in which a deviation rather than the value of the phase is registered, a small admixture of the parasitic modes does not constitute a significant problem. We believe that the MZI can be further optimized or redesigned and that other important PIC components operating on higher-order modes can also be proposed.

\section{Conclusions}

In this work, we have shown that higher-order azimuthally polarized modes can be used in PICs instead of fundamental modes, making it possible to significantly reduce the waveguide separation distance without increasing the interwaveguide crosstalk. Furthermore, we have found that, compared to fundamental modes, the AZ modes can be essentially free of crosstalk in bent waveguide arrays with relatively large radii of curvature. Such bent arrays can be used, e.g., in on-chip LiDAR scanners, waveguide routers, and multiplexers. Since the crosstalk-free separation of the waveguides we consider can be much smaller than those of conventionally used waveguides, the size of a PIC can be reduced by a large factor when using the presented approach. 

We have proposed two ways to selectively excite the AZ modes by coupling light from a thinner single-mode waveguide to a thicker waveguide. One scheme is fabrication-friendly, and the other one provides a nearly unity power conversion efficiency. We have also designed an ultracompact directional coupler and a 3-dB splitter for crosstalk-free waveguides operating on AZ modes. These are the basic elements of photonic chips. Using two 3-dB splitters separated by a pair of crosstalk-free waveguides, we have demonstrated the possibility to create a compact Mach-Zehnder interferometer operating on AZ modes.

We believe that other on-chip components, such as interconnects (fan-in and fan-out, star coupler, crossover, etc.), interferometers (e.g., Mach-Zehnder, Michelson, and multipath interferometer), switches and modulators, logic gates, as well as light sources and receivers, are possible to design for AZ modes. Furthermore, the AZ modes are not the only next-order modes after the fundamental ones that allow one to reduce crosstalk. The other eigenmodes are the TM$_{01}$ radially polarized (RA) modes \cite{mauryaCrosstalkReductionClosely2022}. In principle, all the waveguide components can be designed for RA modes as well, or simultaneously for AZ and RA modes, which would double the number of on-chip information channels.

\section*{Acknowledgments}
The work is part of the Research Council of Finland Flagship Programme, Photonics Research and Innovation (PREIN), decision number 346529. The calculations presented above were performed using computer resources within the Aalto University School of Science “Science-IT” project.

\section*{Data availability}
All data that support the findings of this study are included within the article (and any supplementary files).

\section*{Conflict of Interest}

The authors have no conflicts to disclose.

\bibliography{references.bib}
\bibliographystyle{iopart-num}

\end{document}